# SECULAR EVOLUTION OF THE ECCENTRICITY IN THE PSR 1620−26 TRIPLE SYSTEM

FREDERIC A. RASIO
*Institute for Advanced Study, Olden Lane, Princeton, NJ 08540*

**ABSTRACT**   A simple analytic calculation is presented for the secular evolution of the eccentricities in a hierarchical triple system such as the one containing the millisecond pulsar PSR 1620−26 in M4. If the second companion of PSR 1620−26 is of stellar mass ($m_2 \gtrsim 0.1\,M_\odot$), an eccentricity as large as that observed today for the inner binary ($e_1 \simeq 0.03$) could very well have been induced by secular perturbations. In contrast, such a large eccentricity cannot be induced by a second companion of planetary mass.

## INTRODUCTION

The millisecond pulsar PSR B1620−26 in the globular cluster M4 has a low-mass companion (mass $m_1 \simeq 0.3\,M_\odot$ for a pulsar mass $m_p = 1.35\,M_\odot$) in a nearly circular orbit of period $P_1 = 0.524\,\mathrm{yr}$ (Lyne et al. 1988; McKenna & Lyne 1988). The eccentricity $e_1 = 0.0253$, although small, is several orders of magnitude larger than observed in most other low-mass binary millisecond pulsars (Thorsett, Arzoumanian, & Taylor 1993, hereafter TAT). In addition, the timing data reveal a very large value of the pulse period second derivative $\ddot{P} = -2.3 \times 10^{-27}\,\mathrm{s}^{-1}$ (Backer 1993), indicating the presence of an additional source of acceleration. It is very unlikely that this could be caused by a nearby passing star or by the mean gravitational field of M4. Instead, the most likely explanation is that the pulsar has a second, more distant orbital companion (Backer 1993; Backer, Foster, & Sallmen 1993; TAT). Such hierarchical triple systems are expected to be produced quite easily in globular clusters through dynamical interactions between binaries (Mikkola 1984; Hut 1992; Rasio, Hut, & McMillan 1994).

The present timing data are consistent with a second companion mass anywhere in the range $10^{-3}\,M_\odot \lesssim m_2 \lesssim 1\,M_\odot$ with a corresponding orbital period $10\,\mathrm{yr} \lesssim P_2 \lesssim 10^3\,\mathrm{yr}$ (see Michel 1994, and these proceedings). Thus, in particular, there is a possibility that the second companion might be a Jupiter-like planet (Backer 1993; Sigurdsson 1993; TAT), although it could also be another star (main sequence, white dwarf, or even another neutron star). Planets have been detected in orbit around at least one other millisecond pulsar, PSR B1257+12 (Wolszczan & Frail 1992; Wolszczan 1994).

If PSR B1620−26 is indeed in a triple system, then small perturbations of the (inner) binary pulsar's orbit should be induced by the presence of the second (more distant) companion. Rasio (1994) has argued that the unusually large eccentricity of the binary pulsar could be explained naturally as arising from these perturbations, but only if the second companion is a star and *not* a planet. A completely analytic derivation of the induced eccentricity is given here for various limiting regimes of interest. Other treatments of secular evolution in triple systems, using very different approaches, are given by Mazeh & Shaham (1979) and Bailyn (1987).

**SECULAR PERTURBATION THEORY**

Assume for simplicity that the orbital eccentricities and the relative inclination are low. Standard secular perturbation theory of celestial mechanics (e.g., Brouwer & Clemence 1961; see Dermott & Nicholson 1986 for a useful summary) can then be used to calculate the subsequent evolution of the system over a time $t \gg P_2 > P_1$. The general solution for the eccentricities $e_j(t)$ and longitudes of pericenter $\omega_j(t)$ can be written

$$e_j \sin \omega_j = e_j^{(+)} \sin(g_+ t + \beta_+) + e_j^{(-)} \sin(g_- t + \beta_-), \quad (1)$$

$$e_j \cos \omega_j = e_j^{(+)} \cos(g_+ t + \beta_+) + e_j^{(-)} \cos(g_- t + \beta_-), \quad (2)$$

with $j = 1$ for the inner orbit and $j = 2$ for the outer orbit. This solution can be interpreted geometrically as the sum of two vectors describing circles in the $(e_j \cos \omega_j, e_j \sin \omega_j)$ plane (see Fig. IV of Malhotra, these proceedings). Here $g_\pm$ and $e_j^{(\pm)}$ are the eigenvalues and corresponding eigenvector components of the $2 \times 2$ secular perturbation matrix with components

$$A_{11} = \tfrac{1}{4} n_1 \mu_2 \alpha^2 b_{3/2}^{(1)}(\alpha) = \sigma q \alpha, \quad (3)$$

$$A_{12} = -\tfrac{1}{4} n_1 \mu_2 \alpha^2 b_{3/2}^{(2)}(\alpha) = -\sigma q \alpha \beta, \quad (4)$$

$$A_{21} = -\tfrac{1}{4} n_2 \mu_1 \alpha b_{3/2}^{(2)}(\alpha) = -\sigma \nu \beta, \quad (5)$$

$$A_{22} = \tfrac{1}{4} n_2 \mu_1 \alpha b_{3/2}^{(1)}(\alpha) = \sigma \nu, \quad (6)$$

where $\alpha = a_1/a_2 < 1$ is the ratio of semimajor axes, $n_j = 2\pi/P_j$, $\mu_j = m_j/m_p$, and $\sigma = (1/4) n_1 \mu_1 \alpha b_{3/2}^{(1)}(\alpha)$. The dimensionless functions $b_k^{(l)}(\alpha)$ are Laplace coefficients (Brouwer & Clemence 1961). Following the notations of Rasio et al. (1992), we have introduced the quantities $q = \mu_2/\mu_1$, $\nu = n_2/n_1$, and $\beta = b_{3/2}^{(2)}/b_{3/2}^{(1)}$. The eigenvalues of the matrix $A_{ij}$ can then be written explicitly

$$g_\pm = \frac{\sigma}{2} \left\{ q\alpha + \nu \pm \left[ (q\alpha - \nu)^2 + 4q\alpha\nu\beta^2 \right]^{1/2} \right\}, \quad (7)$$

and the eigenvector components are given by

$$\tilde{e}^{(+)} \equiv \frac{e_2^{(+)}}{e_1^{(+)}} = \frac{-A_{21}}{(A_{22} - g_+)}, \quad (8)$$

$$\tilde{e}^{(-)} \equiv \frac{e_1^{(-)}}{e_2^{(-)}} = \frac{-A_{12}}{(A_{11} - g_-)}, \quad (9)$$

up to a normalization. This normalization, as well as the phases $\beta_+$ and $\beta_-$ appearing in equations (1) and (2), must be determined from an initial condition (i.e., values of $e_j$ and $\omega_j$ at $t = 0$).

Here we assume that the (inner) binary pulsar had a very low eccentricity $e_1(0) \simeq 0$ initially, i.e., at the time it acquired its second companion. One can then set arbitrarily $\omega_1(0) = \omega_2(0) = 0$. It is straightforward to show that the corresponding solution of equations (1) and (2) for $t = 0$ implies $\beta_+ = \beta_- = 0$ and eigenvectors with

$$e_1^{(+)} = \frac{\tilde{e}^{(-)} e_2(0)}{\tilde{e}^{(+)} \tilde{e}^{(-)} - 1}, \qquad e_2^{(-)} = \frac{-e_2(0)}{\tilde{e}^{(+)} \tilde{e}^{(-)} - 1}. \tag{10}$$

Solving equations (1) and (2) in general for $e_1(t)$ then gives

$$e_1(t) = \mathcal{F}(q, \alpha) e_2(0) \left[1 - \cos(gt)\right]^{1/2}, \tag{11}$$

where $g = g_+ - g_-$ and

$$\mathcal{F}(q, \alpha) = \sqrt{2} \left| \frac{\tilde{e}^{(-)}}{1 - \tilde{e}^{(+)} \tilde{e}^{(-)}} \right| \tag{12}$$

(cf. Rasio 1994). Thus the eccentricity of the inner binary returns periodically to its initial value $e_1(0) \simeq 0$, with a maximum amplitude $e_{1,max} = \sqrt{2} \mathcal{F} e_2(0)$ proportional to the eccentricity of the outer orbit.

The analytic solution given by equations (7)–(12) is valid only for low eccentricities and low relative inclination. In general, numerical integrations of the three-body problem must be used (Rasio 1994). One finds that the form of the solution remains generally similar to that of equation (11), with $e_1(t)$ returning to zero periodically. However, both the period and the amplitude of the solution can deviate significantly from their analytic values when $e_2$ or the inclination is very large.

**LIMITING SOLUTIONS FOR $\alpha \ll 1$**

In a strongly hierarchical triple system one can assume that $\alpha \ll 1$. The relevant Laplace coefficients can then be expanded as $b_{3/2}^{(1)} = 3\alpha + \mathcal{O}(\alpha^3)$ and $b_{3/2}^{(2)} = (15/4)\alpha^2 + \mathcal{O}(\alpha^4)$ (Brouwer & Clemence 1961), so that $\beta \simeq (5/4)\alpha$. Using $\nu \simeq \alpha^{3/2}$ we get from equation (7)

$$g_\pm \simeq \frac{\sigma}{2} \left\{ q\alpha + \alpha^{3/2} \pm \left[q^2 \alpha^2 + \alpha^3 - 2q\alpha^{5/2}\right]^{1/2} \right\}. \tag{13}$$

To proceed further, we must distinguish between two different limits, corresponding to $q \gg \alpha^{1/2}$ and $q \ll \alpha^{1/2}$. We refer to these limits as the "stellar" and "planetary" cases, respectively.

In the stellar case one finds that the eigenvalues given by equation (13) can be written

$$g_+ \simeq \sigma q \alpha, \qquad g_- \simeq \sigma \alpha^{3/2} \ll g_+, \tag{14}$$

and the corresponding eigenvectors, calculated using equations (8) and (9), have

$$\tilde{e}^{(+)} \simeq -\frac{5}{4q}\alpha^{3/2}, \qquad \tilde{e}^{(-)} \simeq \frac{5}{4}\alpha. \qquad (15)$$

Equation (12) then gives

$$\mathcal{F}(q,\alpha) \simeq \frac{5}{2\sqrt{2}}\alpha, \qquad g \simeq g_+ \simeq \frac{3}{4}n_1\mu_2\alpha^3. \qquad (16)$$

Remarkably, the maximum induced eccentricity is *independent of the second companion mass* in this limit. However, this maximum eccentricity is reached after a time inversely proportional to $\mu_2$.

For the planetary case the eigenvalues are

$$g_+ \simeq \sigma\alpha^{3/2}, \qquad g_- \simeq \sigma q\alpha \ll g_+, \qquad (17)$$

and the eigenvectors have

$$\tilde{e}^{(+)} \simeq -\frac{4}{5q}\alpha^{-1/2}, \qquad \tilde{e}^{(-)} \simeq \frac{4}{5}\alpha^{-1}, \qquad (18)$$

giving

$$\mathcal{F}(q,\alpha) \simeq \frac{5}{2\sqrt{2}}q\,\alpha^{1/2}, \qquad g \simeq g_+ \simeq \frac{3}{4}n_1\mu_1\alpha^{7/2}. \qquad (19)$$

Here the *period* of the eccentricity variations is independent of $\mu_2$, but the maximum amplitude decreases like the mass ratio $q$.

## APPLICATION TO THE PSR 1620−26 TRIPLE

For PSR 1620−26 we have $\alpha \gtrsim 10^{-2}$ and $q \sim 10^{-3}$ for a Jupiter-size second companion (planetary case) or $q \sim 1$ for a stellar-mass second companion. Using equations (14)–(19) we obtain for the maximum induced eccentricity $e_{1,max} = \sqrt{2}\mathcal{F}e_2$,

$$e_{1,max} \simeq \begin{cases} 3 \times 10^{-2}\,e_2\,(\alpha/10^{-2}) & \text{stellar case} \\ 3 \times 10^{-4}\,e_2\,(q/10^{-3})(\alpha/10^{-2})^{1/2} & \text{planetary case} \end{cases} \qquad (20)$$

Clearly, if the present eccentricity $e_1 = 0.025$ has been induced by secular perturbations, the second companion must be of stellar mass (Rasio 1994).

The period of the eccentricity variations is $P_{e_1} = 2\pi/g$, and we find in the two limiting cases

$$P_{e_1} \simeq \begin{cases} 7 \times 10^5\,\text{yr}\,\mu_2^{-1}\,(\alpha/10^{-2})^{-3} & \text{stellar case} \\ 3 \times 10^7\,\text{yr}\,(\alpha/10^{-2})^{-7/2} & \text{planetary case} \end{cases} \qquad (21)$$

This is certainly much shorter than the age of the pulsar, $t_p \lesssim 10^9$ yr (TAT). It is also shorter than the mean time between close encounters with passing stars in the core of M4 (density $\rho = 10^4\rho_4\,M_\odot\,\text{pc}^{-3}$ and velocity dispersion $\sigma = 5\sigma_5\,\text{km s}^{-1}$), which is $t_c \sim 10^8$ yr $\rho_4^{-1}\sigma_5(a_2/10\,\text{au})^{-1}$.

**ACKNOWLEDGMENTS**

I am grateful to P. Nicholson for many invaluable discussions. This work has been supported by a Hubble Fellowship, funded by NASA through Grant HF-1037.01-92A from the Space Telescope Science Institute, which is operated by AURA, Inc., under contract NAS5-26555.